\documentstyle[prl,aps,floats,epsf]{revtex}
\begin{document}
\draft
\twocolumn[\hsize\textwidth\columnwidth\hsize\csname
@twocolumnfalse\endcsname
\title{Phonon drag in
ballistic quantum wires
}
\author{M. I. Muradov}
\address{Solid State Physics Department, A.F.Ioffe Institute,
194021 Saint Petersburg, Russia}
\date{\today}
\maketitle
\begin{abstract}
The acoustic phonon--mediated drag contribution to
the drag current created in the ballistic transport
regime in a one--dimensional nanowire by phonons generated by a
current--carrying  ballistic channel in
a nearby nanowire is calculated. The threshold of the
phonon--mediated drag current with respect to bias or gate
voltage is predicted.
\end{abstract}
\vskip 2pc ] 
\section{Introduction}
The purpose of the present paper is to study the phonon
contribution to the drag
current in the course of ballistic (collisionless) electron
transport in a nanowire due to a ballistic driving
current in an adjacent parallel nanowire.
Early the possibility of the
Coulomb drag (CD) effect in the ballistic regime in quantum wires has
been demonstrated by Gurevich, Pevzner and
Fenton~\cite{GPF} and has been experimentally
observed by Debray {\it et al.}~\cite{DVR},~\cite{DVR2}.
Using the approach of~\cite{LAN},~\cite{LB} we consider two parallel
ballistic quantum channels that are connected to two thermal
reservoirs, each being in an independent equilibrium state.
As was recently shown, although most of the heat from
a current through the channel is generated in the reservoirs~\cite{G}
part of the heat is generated by the current carrying
nanostructure itself~\cite{TGG} via
emission of phonons. Electrons penetrating into a biased (drive)
wire from the leads are characterized by different chemical
potentials, the situation is nonequilibrium and the phonons are
generated by the drive wire.

There has been done much work on the phonon
contribution to the drag current in a two-dimensional electron
gas situation~\cite{GRAM}. However,
we are not aware of
any work on the phonon drag in ballistic quantum wires.
The electrons in the nearby (drag)
nanowire being initially in an equilibrium absorb the ballistic
phonons
emitted by the drive wire and the phonon drag current is
created. Similar to the CD situation here we encounter
that in the course of backscattering of electrons in the drive wire
the phonons with quasimomentum $\hbar q_z=2p_n$
are generated that in their turn are absorbed by the drag wire.
The contribution to the current of two subbands
$\varepsilon_l+p^2/2m$ in the drive wire and
$\varepsilon_n+p^2/2m$ in the drag wire vanishes unless
$$eV/2\,>\,|\varepsilon_l-\varepsilon_n|$$ again similar to the
CD case~\cite{GM}. However, in the phonon drag we encounter
the threshold $$eV/2\,>\,sp_n.$$ Indeed, the energy-momentum conservation
leads to
$$
\hbar\omega_{\bf q}=\varepsilon_{p+\hbar q_z}-\varepsilon_p
$$
that can be rewritten using $q\cos{\theta_q}=q_z$ as
$$
|\cos{\theta_q}|={2ms\over{|p+\hbar q_z|-p}}\,<\,1
$$
Now taking into account that $|p+\hbar q_z|,p$ should be in the
vicinity of $p_n\pm\,eV/2v_n$ we get the threshold condition $eV/2\,>\,sp_n$.
\section{Phonon spatial distribution}
We assume that the length of the nanowires $L$ is much greater
than the transverse dimensions of the wires. Therefore the
spatial distribution of the emitted by the {\em drive} wire ballistic
(nonequilibrium) phonons with
wave vector ${\bf q}$, $\delta N_{\bf q}({\bf r})$ given
by the stationary Boltzmann equation
\begin{equation}\label{be}
{\bf s}\nabla \delta N_{\bf q}={\cal R}
\end{equation}
where ${\bf s}={\partial\omega_{\bf q}}/\partial{\bf q}$ is the
group sound velocity, $\hbar\omega_{\bf q}$ is the phonon energy, ${\cal R}$ is the collision operator
describing phonon generation,
can be rewritten in the form
\begin{equation}\label{be01}
(\partial_x+\alpha\partial_y) \delta N_{\bf q}(x,y)={{\cal
R}\over{s_x}},\;\;\alpha\equiv\,
{s_y\over{s_x}}=\tan{\varphi_q}.
\end{equation}
We assume that the spatial distribution of the
generated phonons depends only on the transverse coordinates
$x,y$.
The collision operator ${\cal R}$ can be written as
\begin{eqnarray}\label{colop}
{\cal R}={1\over{{\cal
V}}}\sum_{ll'}\int{2Ldk\over{2\pi\hbar}}W_{\bf
q}\left|C_{ll'}({\bf q}_{\perp})\right|^2\nonumber\\
\times\left[
f_{l,k+\hbar q_z}(1-f_{l'k})(N_{\bf q}^{eq}+1)\right.\nonumber\\
\left.-f_{l',k}(1-f_{l,k+\hbar q_z})N_{\bf q}^{eq}
\right]\nonumber\\
\times\delta(\varepsilon_{l,k+\hbar
q_z}-\varepsilon_{l',k}-\hbar\omega_{\bf q}),
\end{eqnarray}
where ${\cal V}$ is the volume of the channel, $C_{ll'}(\bf
q)=<l|e^{i{\bf qr}_{\perp}}|l'>$ is the matrix element for
phonon induced transitions, $W_{\bf q}$ is the
electron--phonon coupling constant that for the deformation
potential interaction is $W_{\bf
q}=\pi\Lambda^2q^2/\rho\omega_{\bf q}$, where $\Lambda$ is the
deformation potential constant, $\rho$ is the mass density.

According to approach in~\cite{LAN},~\cite{LB} we express the distribution
functions in Eq.(\ref{colop}) as the Fermi functions
$f^F(\varepsilon_{lp}-\mu^{(R,L)})$ with shifted chemical
potentials $\mu^{(R,L)}=\mu\pm\,eV/2$, $\mu$ is the quasi Fermi
level that depends on the gate voltage and $V$ is the bias voltage.

What
is concerning the collision operator spatial dependence we assume that it has
nonzero values only within the nanowire.

Restricting ourselves by the case $T=0$ we consider
only the spontaneous term in Eq.(\ref{colop})
since there is no equilibrium phonons $N_{\bf
q}^{eq}=0$.
\begin{eqnarray}\label{spem}
{\cal R}_{sp}={1\over{{\cal
V}}}\sum_{ll'}\int{2Ldk\over{2\pi\hbar}}W_{\bf
q}\left|C_{ll'}({\bf q}_{\perp})\right|^2
\nonumber\\
\times[f_{l,k+\hbar q_z}
\times(1-f_{l'k})-f_{l,k+\hbar q_z}^{eq}(1-f_{l'k}^{eq})]\nonumber\\
\times\delta(\varepsilon_{l,k+\hbar
q_z}-\varepsilon_{l',k}-\hbar\omega_{\bf q})
\end{eqnarray}
Here in Eq.(\ref{spem}) we take into account
explicitly that
phonons are generated by the nonequilibrium electrons.

We assume the channels have uniform cross sections, the
origin of the system of reference being
in the center of the current--carrying (drive) wire, the drag wire being
shifted along the $x$ axis by the distance $D$. Therefore we need
the solution of Eq.(\ref{be01}) only for $s_x\,>0$.

\begin{figure}[htb]
\epsfxsize=2.4in
\epsffile{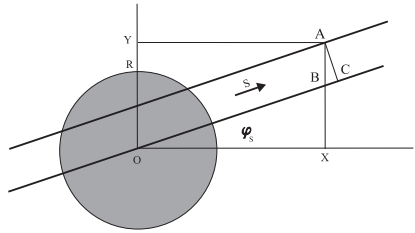}
\caption{Schematic representation of the
cross section of the drive wire (shaded region)
and the direction of phonons' propagation.}
\label{fig1}
\end{figure}

The solution of the Eq.(\ref{be01}) depends on the
cross section geometry of the wire. Assuming, for simplicity, that the
cross section of the wire is a circle (it is worth noting that
the result does not change significantly for other geometries of
the cross section) of the radius $R$ we get
for the coordinates outside the wire cross section
\begin{eqnarray}\label{circle}
\delta N_{\bf q}={2{\cal
R}_{sp}\over{s_x(1+\alpha^2)}}\sqrt{R^2(1+\alpha^2)-(y-\alpha
x)^2} \nonumber\\
\times\Theta[R^2(1+\alpha^2)-(y-\alpha
x)^2].
\end{eqnarray}
The geometrical interpretation of this solution is physically
transparent (see Fig.~\ref{fig1}):
let us draw a line through the center of the cross
section of the wire having the angle $\varphi_s$,
$\tan\varphi_s=\alpha$ with $x$ axis. Now consider a line
parallel to the already drawn one and crossing the cross section
of the wire. The distance $|AC|$ from the point $x,y$ outside the cross
section of the
wire and lying on the second line to the first line is
$|y-\alpha x|\cos{\varphi_s}=|y-\alpha x|/\sqrt{1+\alpha^2}$.
The $\Theta$-function in Eq.(\ref{circle}) states that if this
distance is smaller than the radius R (i.e. the second line does
cross the cross section of the wire) the result is proportional
to the length of the chord cut from the second line by the cross
section, otherwise the result is zero.
The similar
interpretation is valid for the other geometries of the wire
cross section.
\section{Phonon drag contribution to the drag current}
Now in the {\em drag} wire due to an electron-phonon interaction we have
$f_n=f_n^{(F)}+\Delta f_n$ with $\Delta f_n$ satisfying the
equation~\cite{GPF}
\begin{equation}\label{dffe}
v{\partial \Delta f_n\over{\partial z}}=I[f_n],
\end{equation}
where $v=\partial \varepsilon_n/\partial p$ is the electron
velocity, $I[f]$ is the electron--phonon collision term. For
$p>0\,(p<0)$ respectively the solution of this equation is
\begin{equation}\label{dffe01}
\Delta f_n(z)=(z\pm L/2){1\over{v}}I[f_n].
\end{equation}
The boundary condition is $\Delta f[p>0\,(p<0)]=0$ at $z=\mp
L/2$. The current then is given by
\begin{eqnarray}\label{current}
J=e{1\over{\cal V}}\sum_{n,p}\int d{\cal A}v\Delta
f_n\nonumber\\
=e\sum_n\int{dxdy\over{{\cal A}}}\int_0^{\infty}{2Ldp\over{2\pi\hbar}}I[f_n].
\end{eqnarray}
Here the integration is over the cross section ${\cal A}$ of the drag wire.
The electron phonon collision term is
\begin{eqnarray}\label{colterm}
I[f_n]=\sum_{n'}\int{dp'\over{2\pi\hbar}}\int{d{\bf
q}_{\perp}\over{(2\pi)^2}}
W_{\bf q}\left|C_{nn'}({\bf q}_{\perp})\right|^2\nonumber\\
\times\{
\left[
f_{n'p'}(1-f_{np})(N_{p'-p}+1)\right.\nonumber\\
-\left.f_{np}(1-f_{n'p'})N_{p'-p}
\right]\nonumber\\
\times\delta(\varepsilon_{np}-\varepsilon_{n'p'}+\hbar\omega_{p'-p})\nonumber\\
+\left[
f_{n'p'}(1-f_{np})N_{p-p'}\right.\nonumber\\
-\left. f_{np}(1-f_{n'p'})(N_{p-p'}+1)
\right]\nonumber\\
\times\delta(\varepsilon_{np}-\varepsilon_{n'p'}-\hbar\omega_{p-p'})
\}.
\end{eqnarray}
In $N_{\bf q}$ and $\omega_{\bf q}$ we indicate explicitly only
the longitudinal component of quasimomenta of phonons. The first
and second term in Eq.(\ref{colterm}) describes the $N_{p'-p}$ phonon
generation in the electron transition
$\varepsilon_{n'p'}\rightarrow\varepsilon_{np}$ and absorption
of the phonon $N_{p'-p}$ via the electron transition
$\varepsilon_{np}\rightarrow\varepsilon_{n'p'}$ respectively.
The third and forth terms describe absorption of the phonon
$N_{p-p'}$ ($\varepsilon_{n'p'}\rightarrow\varepsilon_{np}$) and
generation ($\varepsilon_{np}\rightarrow\varepsilon_{n'p'}$) of the phonon
$N_{p-p'}$.
For the current induced in the {\em drag} wire  by the phonons emitted
by the {\em drive} wire we get
\begin{eqnarray}\label{curr01}
J=e\sum_{n,n'}\int{dxdy\over{{\cal
A}}}\int_0^{\infty}{2Ldp\over{2\pi\hbar}}
\int_{-\infty}^{\infty}{dp'\over{2\pi\hbar}}
\nonumber\\
\int\,{d{\bf
q}_{\perp}\over{(2\pi)^2}}
W_{\bf q}\left|C_{nn'}({\bf
q}_{\perp})\right|^2[f_{n'p'}-f_{np}] \nonumber\\
\times\left\{\delta
N_{p'-p}\delta(\varepsilon_{np}-\varepsilon_{n'p'}+\hbar\omega_{p'-p})\right.\nonumber\\
+\delta
N_{p-p'}\left.\delta(\varepsilon_{np}-\varepsilon_{n'p'}
-\hbar\omega_{p-p'})\right\}.
\end{eqnarray}
Here the distribution functions $f_{(n,n')}$ are equilibrium
Fermi functions $f^F(\varepsilon_{np}-\mu)$.
Assuming that the angle dependence is involved
only through the phonon distribution we can take an average over
the cross section of the drag wire and integrate over the angles
$\varphi_q$
\begin{eqnarray}
\int{dxdy\over{\pi R^2}}\int{q_{\perp}dq_{\perp}\over{(2\pi)^2}}
d\varphi_q\Delta
N_{\bf q}={2R\over{(2\pi)^2s\pi}}\\
\times\int{qdq_{\perp}}\,{\cal R}_{sp}
\int_0^1\rho d\rho\int_0^{2\pi}d\varphi
\int_{-\pi/2}^{\pi/2}d\varphi_q\nonumber\\
\times\sqrt{1-\left(\rho\sin{\varphi}-{D\over{R}}\sin{\varphi_q}\right)^2}.\nonumber
\end{eqnarray}

Since $D/R\gg\,1$ we see that only small angles contribute to
the integral
$$
{R\over{D}}(\rho\sin{\varphi}-1)<\sin{\varphi_q}\simeq\,
\varphi_q\,<\,{R\over{D}}(\rho\sin{\varphi}+1).
$$
Therefore, the result
$$
\int_{-{R\over{D}}(1-\rho\sin{\varphi})}^{{R\over{D}}(\rho\sin{\varphi}+1)}d\varphi_q
\sqrt{1-\left(\rho\sin{\varphi}-{D\over{R}}{\varphi_q}\right)^2}=
{\pi\over{2}}{R\over{D}}
$$
is proportional to the (solid) angle $R/D$ of the cross
section of the drag wire relative to the drive wire. This factor
simply reflects the fact that phonons emitted by the drive wire
should penetrate the drag wire.
\begin{eqnarray}
\int{dxdy\over{\pi R^2}}\int{q_{\perp}dq_{\perp}\over{(2\pi)^2}}
d\varphi_q\Delta
N_{\bf q}=\nonumber\\
{1\over{4\pi}}{R\over{s}}{R\over{D}}
\int{qdq_{\perp}}\,{\cal R}_{sp}.
\end{eqnarray}
Inserting this expression in the expression for the current and
taking into account energy conservation laws
\begin{eqnarray}\label{encons}
\delta(\varepsilon_{np}-\varepsilon_{n'p'}+\hbar\omega_{p'-p})\nonumber\\
\times\delta(\varepsilon_{lk+p'-p}-\varepsilon_{l'k}-\hbar\omega_{p'-p})=\nonumber\\
={m\over{\hbar s|p-p'|}}\delta\left(k-p-{m\delta\over{p-p'}}\right)\nonumber\\
\times\delta\left(q-{\varepsilon_{n'p'}-\varepsilon_{np}\over{\hbar s}}\right),
\end{eqnarray}

that allows the integration over $k$ we get
\begin{eqnarray}\label{curr02}
J=e{Lm\pi\Lambda^4\over{8D\rho^2(\pi s\hbar)^4}}\sum_{nn'll'}
\int_0^{\infty}dp
\int_{-\infty}^{\infty}dp'\nonumber\\
\int_{|p-p'|/\hbar}^{\infty}
{q^4dq\over{\sqrt{q^2-(p-p'/\hbar)^2}}}
\left|C_{nn'}
({\bf q}_{\perp})\right|^2\nonumber\\
\times\left|C_{ll'}({\bf
q}_{\perp})\right|^2
[f_{n'p'}-f_{np}]{1\over{|p-p'|}}\nonumber\\
\times\{ f_{lp'+m\delta/(p-p')}(1-f_{l'p+m\delta/(p-p')})\nonumber\\
\times\delta\left(q-{\varepsilon_{n'p'}-\varepsilon_{np}\over{\hbar s}}\right)
\nonumber\\
+f_{lp+m\gamma/(p-p')}(1-f_{l'p'+m\gamma/(p-p')})\nonumber\\
\times\delta\left(q-{\varepsilon_{np}-\varepsilon_{n'p'}\over{\hbar s}}\right)\},
\end{eqnarray}
Here $\delta$ and $\gamma$ stand for
$\delta=\varepsilon_{n}+\varepsilon_{l}-\varepsilon_{n'}-\varepsilon_{l'}$
$\gamma=\varepsilon_{n}+\varepsilon_{l'}-\varepsilon_{n'}-\varepsilon_{l}$.
Integrating over $q$ we get
\begin{eqnarray}\label{curr03}
J=e{Lm\Lambda^4\over{8D\rho^2\pi^3(s\hbar)^7}}\sum_{nn'll'}
\int_0^{\infty}dp
\int_{-\infty}^{\infty}dp' \nonumber\\
\times[f_{n'p'}-f_{np}] {P_{nn'll'}(p,p')\over{|p-p'|}}\nonumber\\
\times\left\{ f_{lp'+m\delta/(p-p')}(1-f_{l'p+m\delta/(p-p')})\right.\nonumber\\
\Theta\left(\varepsilon_{n'p'}-\varepsilon_{np}-s|p-p'|\right)\nonumber\\
+f_{lp+m\gamma/(p-p')}(1-f_{l'p'+m\gamma/(p-p')})\nonumber\\
\left.\Theta\left(\varepsilon_{np}-\varepsilon_{n'p'}-s|p-p'|\right)\right\},
\end{eqnarray}
where we introduced notation
\begin{eqnarray}
P_{nn'll'}(p,p')={(\varepsilon_{n'p'}-\varepsilon_{np})^4
\over{\phi_{n,n'}(p,p')}}\nonumber\\
\times\left|C_{nn'}\left({1\over{\hbar s}}
\phi_{n,n'}(p,p')\right)\right|^2\nonumber\\
\times\left|C_{ll'}\left({1\over{\hbar s}}
\phi_{n,n'}(p,p')\right)\right|^2,
\end{eqnarray}
where
\begin{eqnarray}
\phi_{n,n'}(p,p')=\sqrt{(\varepsilon_{n'p'}-\varepsilon_{np})^2-s^2(p-p')^2}.
\end{eqnarray}
Let us consider the case when $n=n'$ and $l=l'$. Then $\delta=\gamma=0$.

\begin{figure}[htb]
\epsfxsize=2.4in
\epsffile{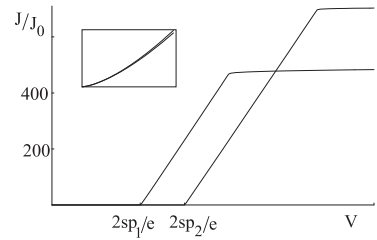}
\caption{Drive voltage dependence of the
phonon contribution to the drag current for different values of
$p_n$.}
\label{fig2}
\end{figure}

Now taking into account that the phonons are emitted by the electrons
having negative initial momentum [the nonequilibrium functions
$f_{l,p}$ are Fermi functions with shifted by $eV/2>0$ chemical potentials,
i.e. $f_{lp}=f^{L}_{lp}$ for $p>0$ and $f_{lp}=f^{R}_{lp}$ for
$p<0$
($f^L=f^F(\varepsilon-[\mu-eV/2])$, $f^R=f^F(\varepsilon-[\mu+eV/2])$)]
and that the integrals vanish unless the Fermi functions $f$ and
$1-f$ under the
integral overlap one
conclude that $p'$ satisfying $-\infty\,<p'<\,-p-2p_s$, $p_s=ms$,
contribute to the current. To avoid further confusion about the
drag current direction note that we consider $eV>0$ and
therefore the first term in the right hand side of
Eq.(\ref{curr03}) survives, otherwise if $eV<0$ only the second
term in this equation would contribute to the current and due to
$f_{np'}-f_{np}$ the current would change the sign.
We
have after transformation $p'\rightarrow\,-p'$
\begin{eqnarray}\label{curr031}
J=e{Lm\Lambda^4\over{8D\rho^2\pi^3(s\hbar)^7}}\sum_{nl}
\int_0^{\infty}dp
\int_{-\infty}^{-p-2p_s}dp'\nonumber\\
\times [f_{np'}-f_{np}]{P_{nl}(p,p')\over{|p-p'|}}
f_{lp'}^{R}(1-f_{lp}^{L})\nonumber\\
=e{Lm\Lambda^4\over{8D\rho^2\pi^3(s\hbar)^7}}\sum_{nl}
\int_0^{\infty}dp
\int_{p+2p_s}^{\infty}dp'\nonumber\\
\times[\Theta(p_n-p')-\Theta(p_n-p)]\nonumber\\
\times{P_{nl}(p,-p')\over{p+p'}}
\Theta(p-p_l^-)\Theta(p_l^+-p'),
\end{eqnarray}
where $$p_n=\sqrt{2m(\mu-\varepsilon_n)},\;\;
p_n^{\pm}=\sqrt{2m(\mu-\varepsilon_n\pm eV/2)}.$$
We get the nonzero result only if
\begin{equation}
p_l^+-p_l^-\,>2p_s,\;\;\;
p_l^-<p_n<p_l^+
\end{equation}
The last inequality is equivalent to
$$eV/2\,>\,|\varepsilon_{n}-\varepsilon_{l}|=|\varepsilon_{nl}|.$$
Assuming $eV/2\ll\,p_n^2/2m$, $|\varepsilon_{nl}|\ll\,p_n^2/2m$
we have nonzero result only if
$$eV/2\,>\,\mbox{max}\{sp_n,|\varepsilon_{nl}|\}.$$
Let us put $\varepsilon_{nl}=0$,
we obtain for $eV/2\,<\,2sp_n$ ($\alpha\equiv\,eV/4sp_n$), i.e. $1/2\,<\alpha\,<1$
\begin{eqnarray}
J=J_0\sum_{n}\left({v_n\over{s}}\right)^2
\int_{1}^{2\alpha}dt(2\alpha-t)T(t)
\end{eqnarray}
and for $eV/2\,>\,2sp_n$ ($\alpha\,>1$)
\begin{eqnarray}
J=J_0\sum_{n}\left({v_n\over{s}}\right)^2
\left\{\int_{1}^{\alpha}dt\,t\right.\nonumber\\+
\left.\int_{\alpha}^{2\alpha}dt(2\alpha-t)\right\}T(t).
\end{eqnarray}
\begin{equation}
J_0=-e{2Lm^5\Lambda^4\over{\pi^3D\rho^2\hbar^7}}
\end{equation}
\begin{eqnarray}
T(t)={t^4
\over{\sqrt{t^2-1}}}
\left|C_{nn}\left({2p_n\over{\hbar}}
\sqrt{t^2-1}\right)\right|^2\nonumber\\
\times\left|C_{ll}\left({2p_n\over{\hbar}}
\sqrt{t^2-1}\right)\right|^2.
\end{eqnarray}
Assuming the following model dependence for $C_{nn}$
\begin{equation}
|C_{nn}(q)|^2={1\over{(1+qR)^2}}
\end{equation}
we plot in Fig.~\ref{fig2} the drag current versus the voltage applied
across the drive
nanowire for different $p_n$ values
$p_1R/\hbar=3.33$ and $p_2R/\hbar=5$.
Near the threshold $\alpha-1/2\ll\,1$ we get, assuming
 $2\alpha-1\ll\,(\hbar/2p_nR)^2$ so that the argument of the function $C_{nn}$
is small and $C_{nn}\sim\,1$
\begin{eqnarray}
J=-e{4\sqrt{2}Lm^5\Lambda^4\over{3D\rho^2\pi^3\hbar^7}}
\left({v_n\over{s}}\right)^2\left({eV\over{2sp_n}}-1\right)^{3/2},\\
1<{eV\over{2sp_n}}(<2)\nonumber
\end{eqnarray}
i.e. at the threshold the current
increases nonlinearly with the applied voltage. This dependence is illustrated
by the small inlet in Fig.~\ref{fig2} since it can be noted only in very small vicinity
near the threshold.

The second derivative of the
current with respect to the bias voltage diverges as
$$
{d^2J\over{dV^2}}\sim {1\over{\sqrt{eV/2sp_n-1}}}
$$
at the threshold for any new subband $n$, this fact can be
instrumental for the experimental investigation.

Assuming the following parameters
$eV/4sp_n\geq\,1$, (for $s\simeq\,3\cdot\,10^5\,\,$ cm/s,
$\hbar/p_n\simeq\,0.5\cdot\,10^{-6}\,\,$ cm this means voltages
$V\sim$ mV), $p_nR/\hbar\simeq\,1$, $L\simeq\,2\,\,\mu$m,
$D\sim\,0.1\cdot\mu$m, $\Lambda\sim\,8\,$ eV, $m=0.07m_0$ we
get the following estimation for the contribution of any subband
to the phonon-mediated drag current
$$
J\sim\,3\cdot\,10^{-12}\;A
$$
\section{Conclusion}
We note two essential differences between the phonon
drag and the Coulomb drag. First, it is the existence of the
threshold $eV/2\,>\,sp_n$
(that can be achieved changing either the bias voltage or the
gate voltage). Second, the weak dependence on the distance $D$
between the centers of the drive and drag wires
rather than the exponential one in the CD case.
On the other hand, note, that the distance dependence of the phonon drag
current in our case is stronger than the distance dependence of
the phonon drag  between two 2DEG layers~\cite{GRAM}.

The coupling coefficient for the piezoelectric coupling
for GaAs having the cubic ${\bf T_d}$ symmetry
can be written as
$$
W_{\bf q}={\pi\over{\rho\omega_{\bf q}}}
\left[{4\pi e\beta\over{\epsilon}}\right]^2
\left(F(\theta_q,\varphi_q)\right)^2,
$$
where $F(\theta_q,\varphi_q)$ is a function of angles.
The estimations for GaAs
($\beta=10^{5}$ SGS units, $\epsilon=12$, $\Lambda=8$ eV, $\rho=5$ g/cm$^3$,
$s=3\times\,10^{5}$ cm/s) shows that the piezoelectric
interaction is the dominating one for frequencies $\omega_{\bf
q}\leq\,10^{11}\;$ s$^{-1}$.

Frequencies of transmitted phonons in the phonon
drag are $\omega_q\,>\,2s/(\hbar/p_n)$, i.e. are greater than
$10^{12}$ s$^{-1}$.
Therefore, the contribution of the piezoelectric coupling is
smaller than the considered deformation potential coupling.

And finally, one can express the drag current also as an integral over the
transferred phonon momentum $q_z$ in an electron-electron interaction.
We get a physically transparent result
\begin{eqnarray}
J=e{1\over{\cal V}^2}\int_{sw}\,{dxdy\over{{\cal A}}}\sum_{n,l,{\bf
q},q_z>0}\left|W_{\bf q}\right|^2
\nonumber\\
\times\left|C_{ll}({\bf
q}_{\perp})\right|^2
{2q\over{sq_{\perp}}}\sqrt{R^2-s^2\left(yq_x/q-xq_y/q\right)^2}\nonumber\\
\times
\left|C_{nn}({\bf q}_{\perp})\right|^2\Theta(eV-\hbar\omega_{\bf q})\chi_b(\varepsilon_n,\varepsilon_n,\hbar\omega_{\bf
q},q_z)\nonumber\\
\times\chi_a(\varepsilon_l-eV/2,\varepsilon_l+eV/2,\hbar\omega_{\bf
q}-eV,q_z)
\end{eqnarray}
where
\begin{eqnarray}\label{vospr}
&\chi&_a(\varepsilon_{\alpha},\varepsilon_{\beta},\hbar\omega,q)=\nonumber\\
&{1\over \pi}&
\Im\sum_{0<p<\hbar q}
{f(\varepsilon_{\alpha{p-\hbar q}})-f(\varepsilon_{\alpha{p}})\over
{\hbar\omega+(\varepsilon_{\alpha{p}})-
(\varepsilon_{\alpha{p-\hbar q}})-i0}}
\end{eqnarray}
is the Lindhard susceptibility
subject to the constraint $0<p<\hbar q$, while in the
$\chi_b$ there is no the second constraint and the sum is over $p>0$.
This form clearly demonstrates that the drag current is a
convolution of the spontaneous polarizations within each quantum
wire.

\acknowledgements

The author is grateful to V. L. Gurevich, V. D. Kagan, V. I.
Kozub, S. V. Gantsevich for discussions. The author is pleased to
acknowledge the support for this work by the Russian National Fund of
Fundamental Research (Grant No~97-02-18286-a).

\end{document}